\documentclass[10pt,letterpaper,twocolumn]{article} 

\usepackage{ol2}
\usepackage[draft]{hyperref}
\usepackage{amsmath}

\begin{document}

\twocolumn[ 

\title{Active feedback of a Fabry-Perot cavity to the emission of a single InAs/GaAs quantum dot}


\author{Michael Metcalfe,$^{1,*}$ Andreas Muller,$^1$, Glenn S. Solomon,$^{1,2}$, and John Lawall, $^{2}$}

\address{
$^1$The Joint Quantum Institute, NIST and University of Maryland
\\
$^2$National Institute of Standards and Technology, Gaithersburg, Maryland 20899 \\
$^*$Corresponding author: metcalfm@nist.gov
}

\begin{abstract}We present a detailed study of the use of Fabry-Perot (FP) cavities for the spectroscopy of single InAs quantum dots (QDs).
We derive optimal cavity characteristics and resolution limits, and measure photoluminescence linewidths as low as $0.9~\mathrm{GHz}$.
By embedding the QDs in a planar cavity, we obtain a sufficiently large signal to actively feed back on the length of the FP to lock to
the emission of a single QD with a stability below 2\:\% of the QD linewidth.  An integration time
of approximately two seconds is found to yield an optimum compromise between shot noise and cavity length fluctuations.\end{abstract}


 ] 

\noindent \section{Introduction}
A quantum dot~(QD) is a structure which confines electrons in all three dimensions and behaves, as a result, like an artificial atom with discrete  energy levels \cite{Michler2001, Petroff2001}.  Self-assembled InAs/GaAs QDs, for example, have optical transition wavelengths in the near infrared, fluorescence linewidths of the order of $1~\mathrm{GHz}$, and observable fine structure.
Their energy levels can be tuned by a variety of externally applied perturbations, including mechanical stress \cite{Seidl2006}, electric fields \cite{Gotoh2000, Vogel2007}, magnetic fields \cite{Bayer2002, Kowalik2007} and optical fields \cite{Jundt2008, Muller08}.  The workhorse for spectroscopic studies of QD fluorescence has long been a grating
spectrometer employing a cooled CCD (charge-coupled device) detector.  These devices provide high sensitivity and large throughput,
but the resolution is in practice limited to about $7~\mathrm{GHz}$.  Thus, they are not able to resolve the radiative linewidth of single QDs.  For this reason, researchers have recently begun to employ scanning Fabry-Perot (FP) cavities \cite{Vogel2007, Muller08} as higher-resolution probes of QD emission.

\begin{figure}[b]
\includegraphics[scale=0.9]{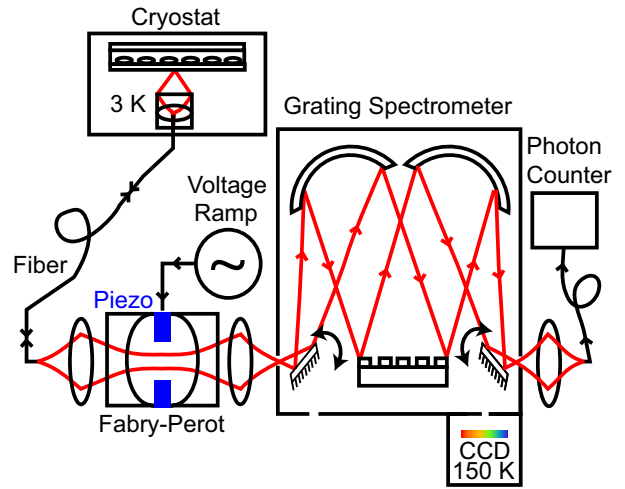}
\caption{\label{MeasSetup} (Color online)
Schematic of the experiment measurement setup.}
\end{figure}

FP cavities are traditionally employed in two distinct manners for the spectroscopy of incident radiation.  In one case, the cavity length and thus resonance frequency is swept. Hence, one can sweep a narrow transmission passband over the frequency range of interest.  In the second case, the cavity resonance frequency is offset from the optical source by approximately a half-linewidth, so that variations in optical frequency are converted into variations in transmitted power.  In both cases, the narrowband nature of the FP transmission limits the throughput relative to that of a grating spectrometer; that is the price to be paid for higher resolution.  This is particularly important for the case of single QD spectroscopy, in which the available signal is very weak.

In this paper, we begin by deriving the FP cavity linewidth that gives the optimum shot-noise limited resolution for a given QD emission spectrum. Then we work out the resulting shot-noise limits to the uncertainty with which measurements of QD emission frequency can be made with a fixed cavity. Next, we illustrate our setup and show representative QD spectra, with widths of approximately 1~GHz.  Finally, we use active feedback to lock the length of a FP cavity to the emission of a single QD.  The feedback loop is implemented in discrete time steps using a digital controller, and
and we have calculated its characteristics in terms of the integration time and gain parameters.  The fluctuations reflect
a compromise between shot noise, which favors longer time steps, and cavity drift, which favors smaller steps.
While the main focus of this work is to treat the QD as a frequency standard to which the FP cavity is locked, the same
ideas should apply to the inverse situation, in which external perturbations are used to lock the emission of a QD to a stable FP cavity.

\begin{figure}[t]
\includegraphics[scale=0.9]{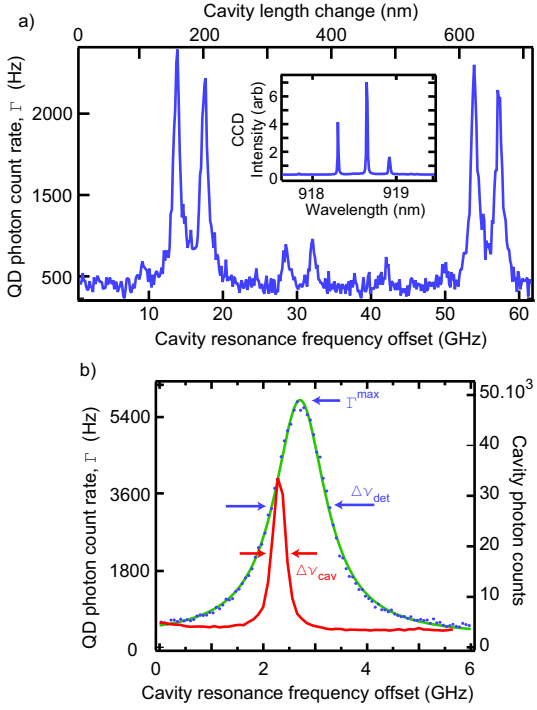}
\caption{\label{DotPeaks} (Color online)
{\bf (a)} Photon count rate vs FP piezo voltage of a QD fluorescence.
The difference between the two sets of (fine-structure split) peaks gives a measure of the cavity free spectral range, $FSR\approx40~\mathrm{GHz}$. The smaller set of split peaks corresponds to the excitation of a higher order mode in the FP cavity. {\bf Inset:} QD spectrum obtained from measuring the QD fluorescence in the spectrometer, without the FP cavity. The tallest peak in this spectrum is selected by the spectrometer and resolved with the FP cavity to give the data shown in (a). {\bf (b)} High resolution spectroscopy of a single QD fluorescence line (dark blue points). The Lorentzian fit (solid green line) gives a full-width at half maximum (FWHM) of $1.2~\mathrm{GHz}$. For comparison, the cavity line is plotted on the same graph (red curve). The cavity has a FWHM of $0.25~\mathrm{GHz}$, implying a QD FWHM of $0.95~\mathrm{GHz}$.}
\end{figure}

\section{Fabry-Perot analysis of QD spectra}
A FP cavity consists of two mirrors separated by a length~$L$, and has a periodic transmission spectrum with peaks separated by
the free spectral range $FSR=c/(2L)$, where $c$ is the speed of light.  In the vicinity of a cavity resonance $\nu_{cav}$, the peaks are approximately Lorentzian,
\begin{equation}
g(\nu)=T\frac{(\frac{\Delta\nu_{cav}}{2})^2}{(\nu-\nu_{cav})^2+(\frac{\Delta\nu_{cav}}{2})^2},
\end{equation}
where $T$ is the resonant power transmission ($T<1$).  The full-width at half maximum (FWHM) is given by
\begin{equation}
\Delta\nu_{cav}=\frac{FSR}{F}.
\end{equation}
where the cavity finesse $F$ is related to the fractional round trip power loss $P_{rt}$ by $F=2\pi/P_{rt}$.
The resonant transmission $T$ depends on the cavity losses and the degree to which
the cavity is impedance-matched \cite{Siegman} and mode-matched\cite{KOGELNIK1966}.

We now consider the transmission of QD fluorescence with a Lorentzian spectral distribution centered at frequency
$\nu_{qd}$ and having FWHM $\Delta\nu_{qd}$:
\begin{equation}
f(\nu)=\Gamma_{QD}\,\frac{\Delta\nu_{qd}}{2\pi}\frac{1}{(\nu-\nu_{qd})^2+(\frac{\Delta\nu_{qd}}{2})^2}.
\end{equation}
This expression is normalized such that the integral over all frequencies gives the total photon
emission rate $\Gamma_{QD}$ sent to the FP cavity.
Considering a particular cavity resonance $\nu_{cav}$ near $\nu_{qd}$, the cavity transmission is the convolution of the QD
fluorescence with the FP response, yielding a Lorentzian in $\nu_{cav}$ centered at frequency $\nu_{qd}$ (see Fig. \ref{DotPeaks}b):
\begin{equation}
\Gamma(\nu_{cav})=\Gamma^{max}
\frac{(\frac{\Delta\nu_{det}}{2})^2}{(\nu_{cav}-\nu_{qd})^2+(\frac{\Delta\nu_{det}}{2})^2},
\label{eqn:convolution}
\end{equation}
with FWHM, $\Delta\nu_{det}$, given by
the sum of the individual linewidths
\begin{equation}
\Delta\nu_{det}=\Delta\nu_{qd}+\Delta\nu_{cav},
\label{eqn:Deltanu}
\end{equation}
and where
\begin{equation}
\Gamma^{max}=\Gamma_{QD} T\frac{\Delta\nu_{cav}}{\Delta\nu_{qd}+\Delta\nu_{cav}}
\label{eqn:Gammamax}
\end{equation}
is the peak detection rate.  Equations~(\ref{eqn:Deltanu}) and~(\ref{eqn:Gammamax})
indicate a tradeoff between throughput (improved by large $\Delta\nu_{cav}$) and resolution
(improved by small $\Delta\nu_{cav}$).

Next, we address the situation in which the FP cavity length is constant
and $\nu_{qd}$ varies in response to some externally applied perturbation.
If a FP transmission resonance is fixed at an offset of approximately a half-linewidth from the center of a QD fluorescence
peak, $|\nu_{cav}-\nu_{qd}|\approx\Delta\nu_{det}/2$ (as illustrated in Fig. \ref{DotPeaks}b),  the throughput is,
to lowest order, linearly related to the frequency offset $\nu_{cav}-\nu_{qd}$.
In the vicinity of this point we now calculate the shot-noise limited uncertainty, $\delta\nu_{qd}$, in a measurement of the QD offset frequency. Changes in $\nu_{qd}$ are related to changes in the detected transmission $\delta\Gamma$ by
\begin{equation}
\delta\nu_{qd}=\frac{1}{d\Gamma/d\nu_{qd}}\delta\Gamma.
\label{eqn: fluctuations}
\end{equation}
Shot noise imposes an uncertainty \cite{Horowitz}
\begin{equation}
\delta\Gamma=\sqrt{\Gamma/\tau},
\label{eqn: shotnoise}
\end{equation}
on a measurement of $\Gamma$ made during an interval $\tau$.
The corresponding limit to the measurement sensitivity is found by substituting equation~(\ref{eqn: shotnoise})
into equation~(\ref{eqn: fluctuations}) and minimizing the resulting expression for $\delta\nu_{qd}$.
One finds that $\delta\nu_{qd}$ is minimized for detunings $\nu_{cav}-\nu_{qd}$ such that
\begin{equation}
\Gamma_{opt}=\frac{2}{3}\Gamma^{max}.
\label{eqn:optcav}
\end{equation}
At this bias point the measurement resolution imposed by shot noise is
\begin{equation}
\delta\nu_{qd}=\frac{3\sqrt{3}}{8\sqrt{T\,\Gamma_{QD} \,\tau}}\frac{(\Delta\nu_{qd}+\Delta\nu_{cav})^{3/2}}{\sqrt{\Delta\nu_{cav}}}.
\label{eqn: shotnoiseres}
\end{equation}
Differentiating with respect to $\Delta\nu_{cav}$, one finds that the optimal shot-noise limited resolution is obtained
for
\begin{equation}
\Delta\nu_{cav}=\frac{\Delta\nu_{qd}}{2},
\label{eqn: optDnu}
\end{equation}
yielding a resolution of
\begin{equation}
\delta\nu_{qd}=\frac{9\sqrt{3}}{16}\frac{\Delta\nu_{qd}}{\sqrt{\Gamma_{max} \,\tau}}.
\label{eqn: ShotNoisednu}
\end{equation}

It has implicitly been assumed in the preceding discussion that the background (dark) count rate,  $\Gamma_{bk}$, is negligible.
If this is not the case, a near-optimal bias point is found by neglecting the shot-noise contribution to the noise and maximizing  $d\Gamma/d\nu_{qd}$. This bias point corresponds to the situation where
\begin{equation}
\Gamma(\nu_{opt})=\frac{3}{4}\Gamma^{max}+\Gamma_{bk}.
\label{eqn:optcavwbk}
\end{equation}
Proceeding as above, the optimal resolution is obtained for
\begin{equation}
\Delta\nu_{cav}=\Delta\nu_{qd},
\label{eqn: optDnuwbk}
\end{equation}
with a shot-noise limited resolution of
\begin{equation}
\delta\nu_{qd}=\frac{4\sqrt{3}}{9}\frac{\Delta\nu_{qd}}{\Gamma_{max}}
\sqrt{\frac{3\Gamma_{max}+4\Gamma_{bk}}{\tau}}.
\label{eqn: ShotNoisednuwbk}
\end{equation}

\section{Experimental implementation}
The FP cavity used in this work employs concave (radius of curvature 25~mm) dielectric mirrors with an average reflectivity
of $98~\%$ over the wavelength interval $875~\mathrm{nm} \textrm{ to } 1000~\mathrm{nm}$, yielding a typical finesse of $F\approx 160$ near
the center of the wavelength range.  It has a length
of 3.8 mm, corresponding to a free spectral range $FSR=40$~GHz and a linewidth $\Delta\nu_{cav}$ of $250~\mathrm{MHz}$.
It was constructed by gluing both mirrors directly to an annular piezoelectric (PZT) actuator.
The cavity is approximately impedance-matched by virtue of the fact that the mirrors have equal
reflectivities.  Mode-matching was accomplished by means of a single achromatic lens between the single-mode fiber
used to collect the QD emission and the FP cavity. With this setup we obtained up to $60~\%$ transmission on the resonance peak of the TEM$_{00}$ mode of the FP.

Our sample consists of InAs QDs embedded in a planar distributed Bragg reflector (DBR) cavity \cite{Solomon2005}. The DBR cavity \cite{Yokoyama1990} is resonant at $920~\mathrm{nm}$ and
consists of a wavelength sized GaAs spacer at $920~\mathrm{nm}$, sandwiched between 15 AlAs/GaAs quarter-wavelength pairs on the bottom and 10 pairs on the top. The directional emission of the DBR cavity increases our collection efficiency, enabling us to resolve QD fluorescence peaks on timescales as low as $1~\mathrm{s}$. We excite the QDs by pumping with a helium-neon laser above the GaAs band gap, resulting in QD emission around $920~\mathrm{nm}$.  The QD density is on the order of $1\, \mu m^{-2}$ and hence we can isolate single QDs by using a tightly focused laser beam.

A diagram of the complete experimental apparatus is given in Fig.~\ref{MeasSetup}.  The sample was maintained
at approximately $4~\mathrm{K}$ in a homemade closed-cycle cryostat.  Excitation light at 633~nm was coupled into the cryostat
via fiber and focused onto the sample by means of a homemade cryogenic microscope
objective (numerical aperture~$\approx 0.4$).  QD fluorescence was captured by the same objective and coupled back into the same fiber.  The microscope objective was scanned over the surface of the sample using cryogenic positioning stages until a strongly emitting QD was located.
Initially, the FP cavity shown in Fig.~\ref{MeasSetup} was not used, and the emission spectrum was measured using a grating spectrometer with a resolution of approximately $8~\mathrm{GHz}$. The spectrometer was used to identify single QD emission lines (inset, Fig. \ref{DotPeaks}(a)). Next, we coupled the QD emission into the scanning FP cavity
and then injected the transmitted light into the grating spectrometer in order to suppress light from all modes of the FP cavity but one;
these modes could potentially transmit light from other QDs located within the laser excitation region.
The output of the spectrometer was directed to a single photon counting module (SPCM).
When scanning the FP cavity, a field-programmable gate array (FPGA) card was used to correlate photon arrival times with PZT voltage,
and histograms of the number of counts vs FP cavity length were acquired.

Typical QD spectra are shown in Fig. \ref{DotPeaks}.
The QD spectrum in Fig. \ref{DotPeaks}a has a fine structure splitting \cite{Gammon96}, contains $10^5$ points and took about $5~\mathrm{minutes}$ to acquire, whereas the QD spectrum in Fig. \ref{DotPeaks}b is a single peak with $10^5$ points and took about $1.5~\mathrm{minutes}$ to acquire.
A Lorentzian fit to Fig. \ref{DotPeaks}b yields a peak width (FWHM) of  $1.2~\mathrm{GHz}$, implying a QD fluorescence spectral width of $950~\mathrm{MHz}$ after deconvolving the broadening from the FP cavity (equation~(\ref{eqn:Deltanu})).  The integration time was sufficiently short that we do not believe that drifts in the FP cavity length significantly contributed to instrumental broadening.
Independent measurements show the radiative lifetime of similar QDs to be longer than $0.8~\mathrm{ns}$, corresponding to a radiative
linewidth below $200~\mathrm{MHz}$. Hence, the emission lines appear to be broadened with respect to the radiative linewidth. This phenomenon is usually attributed to spectral diffusion and is frequently observed \cite{Turck2000, hogele2004, Frantsuzov2008}.

\section{Active stabilization of cavity to QD emission}
We next discuss using the signal transmitted by the FP cavity to lock the cavity resonance to
that of the QD.  This technique could be used in an experiment involving a long data collection run,
during which the cavity may otherwise drift significantly.
The data acquisition can be intermittently halted and the cavity locked to the QD emission
temporarily in order to reestablish the PZT offset voltage corresponding to the QD spectral
feature.  Alternatively, the method may be used to lock the FP cavity to the side of a QD fluorescence peak so
that the cavity functions as a frequency discriminator. The transmission can then be used to infer
the response of the QD emission frequency to an external control parameter, such as stress. Although in many
applications the feedback would be intermittent, in the interest of simplicity, the following discussion assumes the feedback
to be on continuously.

\begin{figure}
\includegraphics[scale=0.9]{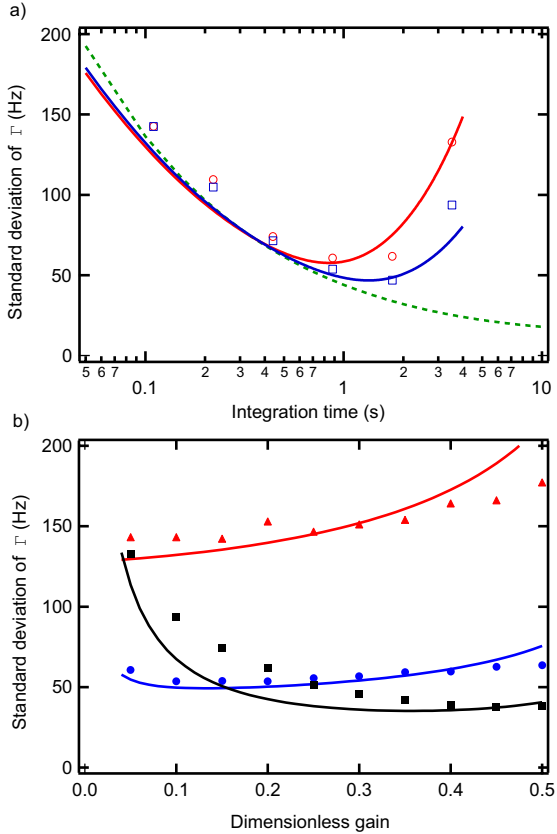}
\caption{\label{SDvstime} (Color online) {\bf (a)} Measured standard deviation of transmitted photon count rate $\Gamma$ as a function of integration time $\tau$,
for $P=0.05$, $I=1$ (circles) and $P=0.1$, $I=1$ (squares).  Solid curves: Corresponding theoretical predictions assuming power spectral density of cavity length fluctuations varies as $\omega^{-2.9}$.  Dashed curve: Theoretical prediction for $P=0.05$, $I=1$ assuming power spectral density
of cavity length fluctuations varies as $\omega^{-1.2}$.  {\bf (b)} Solid points: Measured standard deviation of  $\Gamma$ as a function
of $P$, for $\tau=0.11$~s (triangles), $\tau=0.89$~s (circles), and $\tau=3.56$~s (squares); integral gain $I=1$ in all three cases. }
\end{figure}

In our implementation, the FP cavity is locked to the side of the QD emission peak.  Photons are detected
at the output of the FP cavity and counted for a integration time $\tau$ to infer a rate $\Gamma$.
The difference between the measured count rate and a target count rate $\Gamma^0$ is taken as an error signal,
and sent to a  proportional-integral ($PI$) controller \cite{Franklin01} implemented in software in the
FPGA card.  One of the FPGA analog outputs sends out
a voltage to control the FP cavity length in discrete time steps.
Two sources of fluctuations are considered: Shot
noise, with variance $\Gamma/\tau$, is present with every measurement of $\Gamma$, and the cavity is subject
to environmental disturbances in length $\delta L$ that change its resonance frequency on a slow time scale.
The dynamics are controlled by the integration time $\tau$ as well as the gain parameters $P$ and $I$.
Intuitively, a smaller integration time results in a more agile loop capable
of responding more quickly to changes in the cavity length, at the expense of increased shot noise.

Any feedback loop that forces a lock to the side of a resonance feature has the undesired effect of converting amplitude
noise to frequency noise. In the present case, amplitude variations could come from pump laser power fluctuations or gradual
misalignment of the collection optics and QD. In an independent experiment, we monitored the QD photoluminescence over time without the FP cavity.
We found that for long integration times, where the cavity drift noise dominates the shot noise, the frequency noise that one would infer from the observed QD amplitude fluctuations was small compared to the cavity drifts.
If laser power fluctuations are significant, one could split off some of the reflected pump power from the QD signal for normalization.
Alternatively, one could render the feedback first-order insensitive to changes in signal amplitude by dithering the cavity,
employing phase-sensitive
detection, and locking the transmission to the center of the QD resonance \cite{Baer80}.  While such an implementation would be
straightforward with our software-based controller, we limit the discussion to the side-lock case for simplicity.

The dynamics of the feedback loop are calculated in the Appendix, where the conditions for
stability are derived, as well as the fluctuations expected in response to measurement shot noise
and cavity drifts.  Assuming variations in the cavity length are described by a power spectral density (PSD)
\begin{equation}
S_{\delta L}(\omega)=a\omega^{-b},
\label{eqn:PSD}
\end{equation}
where $\omega$ is the angular frequency and $a$ and $b$ are constants,
the variance in the steady-state photon count rate is found to be
\begin{equation}
\sigma^2_{\Gamma}=\frac{\Gamma^0}{\tau}\left\{1+\frac{P^2}{\pi}J(P,I) \right\} +\frac{a\beta^2}{\pi}\tau^{b-1}Q(b,P,I)
\label{eqn:theory}
\end{equation}
Here $\beta$ is the response of the cavity transmission to changes in its length, $P$ and $I$ are the proportional and integral gains,
and $J(P,I)$ and $Q(b,P,I)$ are integral functions whose explicit forms are given in the Appendix.
The first term decreases with integration time~$\tau$, and reflects the shot noise.  The second term increases with~$\tau$, and reflects
cavity drifts.

We measured the stability of the cavity used in
the measurements described earlier by monitoring the transmission of a frequency-stabilized laser.  The FP cavity was
initially offset by about a half-linewidth from the laser and the transmission was monitored to infer
changes in the cavity length.  Subsequent spectral analysis showed that the data were well described by a power spectral density
\begin{equation}
S_{\delta L}(\omega)\propto\omega^{-1.2}\,{\rm nm^2/Hz}
\label{eqn:Snuexpt}
\end{equation}
for frequencies above 3~mHz.

Next, we implemented the feedback loop with the same FP cavity, using a QD with
a peak count rate of $\Gamma^{max}=3.5~\mathrm{kHz}$ and a deconvolved linewidth $\Delta\nu_{qd}\approx 1.3$~GHz, forcing a nominal
count rate of $\Gamma^0=1.8~\mathrm{kHz}$.
The standard deviation of the steady-state fluctuations are shown in Fig.~\ref{SDvstime}a as a
function of integration time $\tau$ for two
different sets of gain parameters P and I.  The data show the fluctuations to be dominated by shot noise at
small integration times, and cavity length drifts at longer times.
Also shown (dashed line) is the theoretical result given by equation~(\ref{eqn:theory}) and taking
equation~(\ref{eqn:Snuexpt}) to describe the cavity length fluctuations.
It is apparent that the optimal loop performance for these gain parameters is obtained for integration times of approximately $\tau=2\,s$,
whereas the theoretical prediction shows a minimum at approximately $\tau=50\,s$.
We interpret this phenomenon as arising from the fact
that under feedback conditions, the cavity is subject to additional perturbations (primarily thermal)
associated with the fluctuating PZT drive, particularly at very low frequencies.  In fact, we get reasonable agreement
with equation~(\ref{eqn:theory}) if we take $S_{\delta L}(\omega)\propto\omega^{-2.9}\,{\rm nm^2/Hz}$ as the PSD of the cavity while feedback is operational (solid curves in Fig.~\ref{SDvstime}).  We have verified this empirical PSD for a variety of sets of gain parameters as shown in Fig.~\ref{SDvstime}b.
The best case shown in Fig.~\ref{SDvstime}a had a standard deviation ~$\delta\Gamma\approx40\,\mathrm{Hz}$,
corresponding via equation~(\ref{eqn: fluctuations}) to fluctuations in the offset of the cavity resonance from the center
of the QD fluorescence peak of approximately 19~MHz, approximately 1.5\:\% of the QD linewidth.  An optimal choice of $P$, $I$, and $\tau$ should
allow the fluctuations to be reduced to about 13~MHz.  Further improvement could be achieved by using a cavity with a somewhat
larger linewidth, 650~MHz$<\Delta\nu_{cav}<$1.3~GHz (equations (\ref{eqn: optDnu}) and (\ref{eqn:optcavwbk})).

An interesting alternative approach would be to construct the cavity from a material with a high thermal conductivity, such as copper, and actuate its length by means of temperature.  A high level of thermal stability could be ensured by putting the cavity in a temperature-controlled environment, and the high thermal conductivity of the copper should mitigate noise of the sort induced by the fluctuating PZT.

\section{Conclusion}
We have described a hybrid spectrometer for use in QD spectroscopy employing a FP interferometer for high resolution, in conjunction with a grating spectrometer to resolve the ambiguity associated with the periodic FP spectrum.  For optimal noise-limited resolution, we show that the FP cavity should be constructed to have a FWHM linewidth approximately equal to the QD linewidth. With this hybrid spectrometer we have measured QDs with linewidths as low as $0.9~\mathrm{GHz}$ and can easily resolve the QD fine structure. With an appropriate cavity it should be possible to use this technique to resolve the frequency fluctuations contributing to the spectral diffusion, if they are slow enough \cite{hogele2004}. For illustrative purposes, we can apply equation (\ref{eqn:Gammamax}) to the QD and cavity used in the previous section to find a total captured QD fluorescence rate of $\Gamma_{QD} = 43.4~\mathrm{kHz}$.  By biasing at the optimal point on a cavity with transmission $T=1$ and a linewidth $\Delta\nu_{cav}=3~\mathrm{GHz}$, one finds from equation (\ref{eqn: shotnoiseres}) a shot-noise limited resolution of approximately $10~\mathrm{MHz}$ in one second of integration time, largely independent of $\Delta\nu_{qd}$ as long as $\Delta\nu_{qd}<<\Delta\nu_{cav}$.  Assuming the QD to have a natural linewidth $\Delta\nu_{qd}$ of $160~\mathrm{MHz}$, corresponding to radiative decay with a lifetime of $1~\mathrm{ns}$, the measurement resolution is equal to the natural linewidth in an integration time of $4~\mathrm{ms}$.

If the FP cavity is to be used as a probe of the QD fluorescence frequency as a function of an external perturbation, or if the FP cavity is to be used as a sensitive spectroscopic probe for long data collection times, it can be useful to use feedback to lock the cavity to the side of a transmission resonance.  We have achieved a robust lock by measuring the rate at which photons are transmitted by the cavity and implementing a software PI controller to act on the cavity length so as to maintain the cavity at a fixed offset frequency.  An optimal integration time is found to be approximately two seconds, reflecting a tradeoff between measurement shot noise and cavity length fluctuations.
Finally, this work paves the way for the inverse experiment, in which the QD frequency is locked to a stabilized FP cavity using the Stark \cite{Gotoh2000, Vogel2007} or Zeeman \cite{Bayer2002, Kowalik2007} effects.  A particularly interesting application would be to lock the frequencies of two distinct QDs to the same FP peak, so as to create indistinguishable photons from distinct single-photon sources.

\section{Acknowledgments}
The authors would like to thank S. Carr, A. Chijioke, E. B. Flagg and W. Fang for both valuable discussions and technical help throughout this work.

\section{Appendix}
The feedback loop is described by the difference equation
\begin{equation}
{\bf \Gamma}_{n+1}=\Gamma_0+K_{PZT} \,\beta\, P'\left\{\epsilon_n+I\sum_{m=0}^n \epsilon_m \right\} + \beta\,{\bf \delta L}_n
\label{eqn:diff1}
\end{equation}
where we denote by ${\bf \epsilon}_n$ the
``error'' fed to the PID controller at loop cycle $n$:
\begin{equation}
{\bf \epsilon}_n={\bf \Gamma}_n+{\bf \delta\Gamma}_n-\Gamma^0
\end{equation}
Here $K_{PZT}$ is the response of the PZT (m/V), $\beta$ is the (linearized) response of the cavity photon
transmission rate to changes in cavity length (Hz/m), and $P'$ (V/Hz) and $I$ are chosen in software.
The quantities ${\bf \delta\Gamma}_n$ and ${\bf \delta L}_n$ represent the measurement
shot noise and stochastic cavity length fluctuations, respectively.

Subtracting ${\bf \Gamma}_{n}$ from ${\bf \Gamma}_{n+1}$, we obtain
a second-order difference equation for ${\bf \Gamma}_{n}$ with constant coefficients and a stochastic
forcing function:
\begin{eqnarray}
{\bf\Gamma}_{n+1}-\sigma{\bf\Gamma}_n-P{\bf\Gamma}_{n-1}&=& P\,I\,\Gamma^0 \nonumber \\
&&-P\left\{(1+I){\bf \delta\Gamma}_n-{\bf \delta\Gamma}_{n-1} \right\} \nonumber \\
&& +\beta\,({\bf \delta L}_{n}-{\bf \delta L}_{n-1})
\label{eqn: diffeq2}
\end{eqnarray}
where we have defined
\begin{eqnarray}
P&=&-\beta\, K_{PZT} \,P' \\
\sigma&=&1-P(1+I);
\end{eqnarray}
thus $P$ is the (dimensionless) proportional gain all the way around the loop.
Taking the z-transform\cite{Papoulis}, we obtain
\begin{eqnarray}
\Gamma(z)&=&\frac{P\,Iz^2}{(z-1)(z^2-\sigma\,z-P)}\Gamma^0 \\
&&+\frac{P[1-(1+I)z]}{z^2-\sigma\,z-P}\delta\Gamma(z) \\
&&+\frac{\beta(z-1)}{z^2-\sigma\,z-P}\delta L(z)
\end{eqnarray}
The condition for stability is that the poles of the characteristic equation
\begin{equation}
z^2-\sigma\,z-P = 0
\end{equation}
lie within the unit circle $|z|=1$.
After some algebra, one finds that the necessary conditions for stability are
\begin{eqnarray}
0&<&P<1 \\
0&<&I<2/P-2
\end{eqnarray}
Thus both $P$ and $I$ are limited; too much gain will cause instability.

In steady state, the discrete final-value theorem\cite{Franklin01} asserts that
\begin{eqnarray}
\lim_{n\rightarrow\infty}{E\{\bf \Gamma}_{n}\}&=&\lim_{z\rightarrow 1}(1-z^{-1})\Gamma(z) \\
&=&\Gamma^0,
\end{eqnarray}
as desired, provided that $I\ne 0$ and $P\ne 0$.
The variance of ${\bf \Gamma}_n$ is then given by\cite{Papoulis}
\begin{equation}
\sigma^2_{\Gamma}=\frac{1}{2\pi}\int_{-\pi}^\pi S_{\Gamma}^D(\omega)\,d\omega
\end{equation}
where we will express the power spectral density (PSD) $S_{\Gamma}^D(\omega)$ of the discrete
process (indicated by the superscript $D$) ${\bf \Gamma_n}-\Gamma^0$ as the sum of
independent contributions corresponding to
the shot noise at each measurement and the effects of the drive terms ${\bf \delta\Gamma}_n$ and ${\bf \delta L}_n$ in
equation~(\ref{eqn: diffeq2}).

Taking the measurement shot noise to correspond to a stationary white noise process
with zero mean and variance equal to $\Gamma^0/\tau$, the corresponding PSD is simply\cite{Papoulis}
\begin{equation}
S_{\delta\Gamma}^D(\omega)=\Gamma^0/\tau.
\end{equation}
The effect of the shot noise ${\bf \delta\Gamma}$ on the count rate ${\bf \Gamma}$ within the feedback loop is governed
by the transfer function
\begin{equation}
{\bf H}_{\delta\Gamma}(z)=\frac{P[1-(1+I)z]}{z^2-\sigma\,z-P},
\end{equation}
with the corresponding power spectrum $S_{\Gamma}^{(1)}(\omega)$ related to the power spectrum $S_{\delta\Gamma}^D(\omega)$ by\cite{Papoulis}
\begin{equation}
S_{\Gamma}^{(1)}(\omega)=S_{\delta\Gamma}^D(\omega)|{\bf H}_{\delta\Gamma}(e^{i\omega})|^2
\end{equation}
Similarly, the effect of the cavity length fluctuations ${\bf \delta L}_n$ on the count rate ${\bf \Gamma}_n$ within the feedback loop are governed
by
\begin{equation}
{\bf H}_{\delta L}(z)=\frac{\beta(z-1)}{z^2-\sigma\,z-P}
\end{equation}
with the corresponding power spectrum $S_{\Gamma}^{(2)}(\omega)$ given by
\begin{equation}
S_{\Gamma}^{(2)}(\omega)=S_{\delta L}^D(\omega)|{\bf H}_{\delta L}(e^{i\omega})|^2 \\
\label{eqn:Scavity}
\end{equation}
In fact, the cavity length varies during the integration
time $\tau$, so that $S_{\delta L}^D(\omega)$ corresponds to the moving average, sampled at times $n\tau$, of a continuous stochastic process with power spectrum $S_{\delta L}(\omega)$.  The power spectrum of the discrete process ${\bf \delta L}_n$ is given by
\cite{Papoulis}
\[
S_{\delta L}^D(\omega)=\frac{1}{\tau}\sum_{n=-\infty}^\infty {\rm sinc}^2(\frac{\omega+2\pi n}{2})S_{\delta L}(\frac{\omega+2\pi n}{\tau})
\]
where the term containing the sinc function arises from the moving average and the
terms in the sum with $n\ne0$ reflect aliasing from undersampling.

For simplicity, we now neglect the terms arising from aliasing, and in addition assume a
form
\begin{equation}
S_{\delta L}(\omega)=a|\omega|^{-b}
\end{equation}
for the power spectral distribution of the (continuous) cavity length fluctuations.
Combining the previous results, the variance of ${\bf \Gamma}_n$ can then be expressed as
\begin{equation}
\sigma^2_{\Gamma}=\frac{1}{\pi}\int_0^\pi S_{\delta\Gamma}^D(\omega)+S_{\Gamma}^{(1)}(\omega)+S_{\Gamma}^{(2)}(\omega)\,d\omega \nonumber
\label{eqn:final}
\end{equation}
\begin{equation}
=\frac{\Gamma^0}{\tau}\left\{1+\frac{P^2}{\pi}J(P,I) \right\} +\frac{a\beta^2}{\pi}\tau^{b-1}Q(b,P,I)
\label{eqn:final}
\end{equation}
where

\begin{equation}
J(P,I)=
\end{equation}
\begin{equation}
\int_0^{\pi}\frac{2+2I+I^2-2(1+I)\cos(\omega)}{1+P^2+\sigma^2-2\sigma(1+P) \cos(\omega)+2P \cos(2\omega)} \,d\omega,
\end{equation}
\begin{equation}
Q(b,P,I)=
\end{equation}
\begin{equation}
\int_{0}^{\pi} \frac{2[1-\cos(\omega)]\,{\rm sinc}^2(\omega/2)\,\omega^{-b}}{1+P^2+\sigma^2-2\sigma(1-P) \cos(\omega)-2P \cos(2\omega)} \,d\omega
\end{equation}

$J(P,I)$ can be evaluated analytically.  The integral $Q(b,P,I)$ is finite provided $b<3$.

The first term in equation~(\ref{eqn:final}), reflecting the shot noise, can be made smaller by choosing a longer integration time $\tau$ and smaller values
$P$ and $I$ for the feedback.  The second term, reflecting cavity drifts, can be made smaller by choosing a smaller
integration time $\tau$.  It diverges for $I$ approaching either zero or the maximum allowed value $2/P-2$.



\end{document}